\documentclass[sigconf]{acmart} 

\newcommand{\final}{1}

\usepackage{color}
\usepackage{ifthen}
\usepackage{float}
\usepackage{alltt}
\usepackage{amsmath}
\usepackage{amsthm}
\usepackage{newlfont} 
\usepackage{floatflt}
\usepackage{wrapfig}
\usepackage{fixltx2e}
\usepackage{subfig} 
\usepackage{multirow}
\usepackage{booktabs}
\usepackage{cleveref}
\usepackage{algorithmic}
\usepackage{mathtools, cuted}
\usepackage{CJKutf8} 
\usepackage{enumitem}
\usepackage[ruled]{algorithm2e} 
\usepackage{layouts}
\usepackage{pifont}

\setcitestyle{square}
\def\textrightarrow{$\rightarrow$}

\SetAlFnt{\small}
\SetAlCapFnt{\small}
\SetAlCapNameFnt{\small}
\SetAlCapHSkip{0pt}
\IncMargin{-\parindent}

\let\oldcaption\caption
\renewcommand{\caption}[2][]{\oldcaption[#1]{{\em #1} #2}}

\definecolor{SithColor}{rgb}{0.7,0,0} 
\newcommand{\liyi}[1]{{\color{SithColor} Li-Yi: #1 $\qed$}}
\definecolor{ConsularColor}{rgb}{0,0.4,0} 
\definecolor{GuardianColor}{rgb}{0,0,0.8} 
\newcommand{\jiaju}[1]{{\color{GuardianColor} Jiaju: #1 $\qed$}}
\newcommand{\vshin}[1]{{\color{ConsularColor} Jeanette: #1 $\qed$}} 
\newcommand{\jonelson}[1]{{\color{ConsularColor} Erin: #1 $\qed$}} 
\newcommand{\rubaiat}[1]{{\color{ConsularColor} Rubaiat: #1 $\qed$}} 
\newcommand{\maneesh}[1]{{\color{red} MA: #1}} 
\newcommand{\anyi}[1]{{\color{orange} AR: #1}} 
\newcommand{\warning}[1]{{\it\color{red} #1}}

\newcommand{\note}[1]{{\it\color{blue} #1}}
\newcommand{\nothing}[1]{}

\newcommand{\camready}[1]{{\color{GuardianColor}#1}}
\renewcommand{\camready}[1]{{#1}}

\definecolor{AudioColor}{rgb}{0.56,0.34,0.62}

\definecolor{NewColor}{rgb}{0.9,0.4,0}

\definecolor{DeleteColor}{rgb}{0.1,0.6,1.0}
\newcommand{\delete}[1]{{\color{DeleteColor} #1}}
\definecolor{MoveColor}{rgb}{0.5,0.1,0.5}

\definecolor{figred}{rgb}{1,0,0}
\definecolor{figgreen}{rgb}{0,0.6,0}
\definecolor{figblue}{rgb}{0,0,1}
\definecolor{figpink}{rgb}{1,0.63,0.63}

\ifthenelse{\isundefined{\hili}}
{

\renewcommand{\delete}[1]{}

}
{
\ifthenelse{\equal{\hili}{0}}
{

\renewcommand{\delete}[1]{}

}
{}
}

\ifthenelse{\equal{\final}{1}}
{
\renewcommand{\liyi}[1]{}
\renewcommand{\jiaju}[1]{}
\renewcommand{\rubaiat}[1]{}
\renewcommand{\vshin}[1]{}
\renewcommand{\jonelson}[1]{}
\renewcommand{\maneesh}[1]{}
\renewcommand{\anyi}[1]{}

\renewcommand{\warning}[1]{}
\renewcommand{\note}[1]{}

}

\newcommand{\pseudocode}{Algorithm}
\floatstyle{plain}
\newfloat{algorithm}{tbhp}{lop}
\floatname{algorithm}{\pseudocode}

\newcommand{\filename}[1]{\url{#1}}
\newcommand{\foldername}[1]{\url{#1}}

\let\oldparagraph\paragraph
\iffalse

\renewcommand{\paragraph}[1]{\oldparagraph{\textbf{#1}.}} 
\else

\renewcommand{\paragraph}[1]{\oldparagraph{{#1}.}}
\fi

\ifdefined\email
\else
\newcommand{\email}[1]{\url{#1}}
\fi

\if@twocolumn

\fi

\newcommand{\adobeResearch}{Adobe Research}
\newcommand{\stanford}{Stanford University}
\newcommand{\roblox}{Roblox}

\newcommand{\textkernel}{k}

\newcommand{\dg}{\textbf{DG}}
\newcommand{\dgPhrase}{\textbf{\dg1}}
\newcommand{\dgLinebreak}{\textbf{\dg2}}
\newcommand{\dgHighlight}{\textbf{\dg3}}
\newcommand{\dgContrast}{\textbf{\dg4}}
\newcommand{\dgTiming}{\textbf{\dg5}}
\newcommand{\dgFocus}{\textbf{\dg6}}
\newcommand{\dgSpatial}{\textbf{\dg7}}

\newcommand{\energyPrev}{E_{\text{prv}}}
\newcommand{\energyContrast}{E_{\text{cnt}}}

\newcommand{\energyVisualCenter}{E_{\text{fcs}}}
\newcommand{\energyOverlap}{E_{\text{fgd}}}

\newcommand{\spannedFrames}{\mathcal{F}}
\newcommand{\focusMask}{s_{\text{fcs}}}
\newcommand{\foregroundMask}{s_{\text{fgd}}}
\newcommand{\avgFocusMask}{\bar{s}_{\text{fcs}}}
\newcommand{\avgForegroundMask}{\bar{s}_{\text{fgd}}}

\newcommand{\resultsSite}{\url{https://hci.stanford.edu/research/lyricvideo/}}

\setcopyright{acmlicensed}
\copyrightyear{2023}
\acmYear{2023}
\acmConference[UIST '23]{The 36th Annual ACM Symposium on User Interface Software and Technology}{October 29-November 1, 2023}{San Francisco, CA, USA}
\acmBooktitle{The 36th Annual ACM Symposium on User Interface Software and Technology (UIST '23), October 29-November 1, 2023, San Francisco, CA, USA}
\acmPrice{15.00}
\acmDOI{10.1145/3586183.3606757}
\acmISBN{979-8-4007-0132-0/23/10}

\begin{document}

\title{Automated Conversion of Music Videos into Lyric Videos}

\author{Jiaju Ma}
\affiliation{
    \institution{\stanford}
    \country{}
}
\author{Anyi Rao}
\affiliation{
    \institution{\stanford}
    \country{}
}

\author{Li-Yi Wei}
\affiliation{
    \institution{\adobeResearch}
    \country{}
}

\author{Rubaiat Habib Kazi}
\affiliation{
    \institution{\adobeResearch}
    \country{}
}
\author{Valentina Shin}
\affiliation{
    \institution{\adobeResearch}
    \country{}
}


\author{Maneesh Agrawala}
\affiliation{
    \institution{\stanford}
    \country{}
}
\affiliation{
    \institution{\roblox}
    \country{}
}



\begin{abstract}
Musicians and fans often produce lyric videos, a form of music videos that showcase the song's lyrics, for their favorite songs.
However, making such videos can be challenging and time-consuming as the lyrics need to be added in synchrony and visual harmony with the video.
Informed by prior work and close examination of existing lyric videos, we propose a set of design guidelines to help creators make such videos.
Our guidelines ensure the readability of the lyric text while maintaining a unified focus of attention.
We instantiate these guidelines in a fully automated pipeline that converts an input music video into a lyric video.
We demonstrate the robustness of our pipeline by generating lyric videos from a diverse range of input sources.
A user study shows that lyric videos generated by our pipeline are effective in maintaining text readability and unifying the focus of attention.
\end{abstract}




\begin{teaserfigure}
  \centering
  \includegraphics[width=\textwidth]{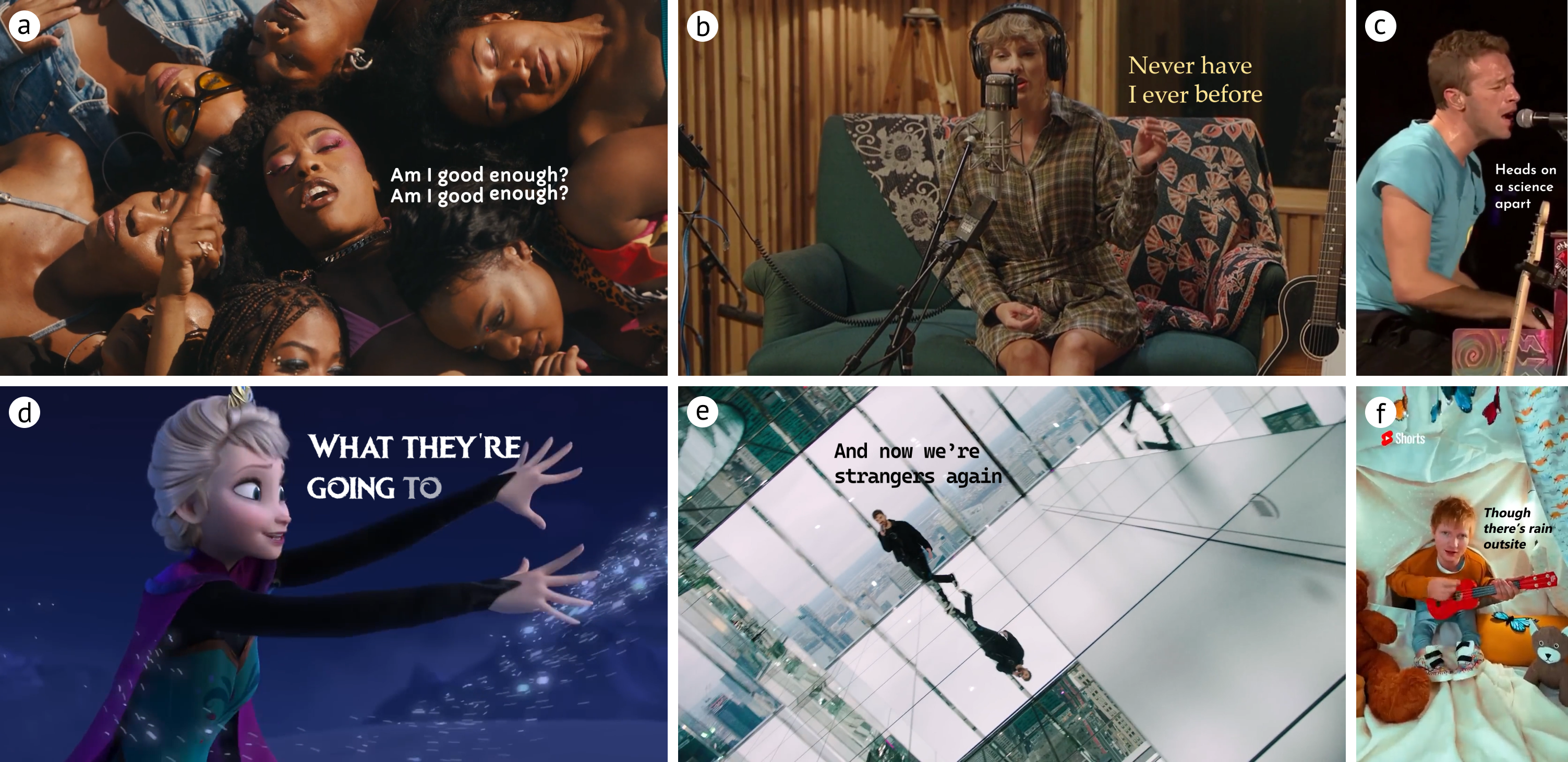}
 \caption{We propose a set of design guidelines for adding lyrics to music videos in a manner that ensures text readability and unifies the viewer's focus of attention.
  We further implement a fully automated pipeline that instantiates these guidelines to convert an input music video into a lyric video.
  The results shown above demonstrate that our pipeline is able to generate lyric videos from a wide variety of inputs.
  \camready{%
  \emph{%
  Video source: 
  (a) Selfish Soul by Sudan Archives~\cite{Sudan:2022:SS}, 
  (b) August by Taylor Swift \textcopyright Taylor Swift~\cite{swift:2021:A},
  (c) The Scientist by Coldplay~\cite{Coldplay:2022:ts}, 
  (d) Let It Go by Idina Menzel \textcopyright 2013 Hollywood Records, Inc.~\cite{Menzel:2013:LIG},
  (e) iPad by Chainsmokers~\cite{Chainsmokers:2022:I}, 
  (f) Sandman by Ed Sheeran~\cite{Sheeran:2022:S}}.
  }
  (a, b, d, e) are in landscape format while (c, f) are in portrait format.
 }
 \label{fig:teaser}
\end{teaserfigure}

\maketitle

\section{Introduction}
%
%


A lyric video is a music video that displays the lyrics with the video imagery.
The history of lyric videos can be traced back to 1965, when Bob Dylan released a video for his song \textit{Subterranean Homesick Blues} in which he flips through a pile of cards with words from the song written on them~\cite{Dylan:1965}.
Following his lead, artists start to release lyric videos, such as Prince~\cite{Prince:1987}, Katy Perry~\cite{Perry:2013}, Taylor Swift~\cite{Swift:2022}, and many more. 
%
Other than featuring the music, lyric videos have many other use cases like lip sync and karaoke. 
Moreover, the presence of text allows the video content to be consumed without audio, either in noisy environments or converted to other media formats like images and GIFs.
One can find many examples of this on platforms like Pinterest and Reddit~\cite{Reddit:HQG}.
%
For these reasons, music fans often make lyric videos themselves by adding animated lyric text to existing music videos, giving birth to YouTube channels like HQG Studios~\cite{YouTube:HQG} whose videos have gathered millions of views.
Social media apps like Instagram and TikTok also have functionalities to help users display song lyrics in their videos.


However, making a lyric video from a music video remains challenging and time-consuming as it requires delicate coordination of audio, visual, and text content~\camready{\cite{Kato:2015:TLI, Vy:2008:EST, Zdenek:2015:RSC}}.
%
First, the creator needs to ensure the readability of the lyric text. 
This entails segmenting the body of text into phrases to be shown in the video sequentially and deliberately adding line breaks when the text is long.
Second, since the added text requires the viewer's attention to read and process, it needs to be in synchrony and visual harmony with the song and video to minimize distractions and unify the viewer's focus.
Achieving these goals requires synchronizing the lyrics to the song and coordinating the text's placement with the video imagery.


Prior work has proposed automated solutions to add text to videos in other forms like subtitles~\cite{Hu:2015:SFV}, kinetic typography~\cite{Kato:2015:TLI, Wang:2017:VVS}, and data visualizations~\cite{Tang:2022:SS}. 
However, these works only partially considered the text readability and the coordination of text, audio, and video, 
but these intertwined challenges need to be taken into account holistically.
%
For example, changing the text of a lyric phrase affects when it should appear temporally, as well as its optimal position within the video frame since its size may also change. 




To help creators make lyric videos that ensure good readability of the text and maintain the viewer's focus of attention, we propose a set of design guidelines formulated
by analyzing guidelines and popular lyric videos.
%
To further assist the creators and validate our proposed design guidelines, we implement a fully automated pipeline that instantiates these design guidelines to convert an input music video to a lyric video.
The user can optionally specify the font, color, size, and animation of the lyric texts and adjust algorithm parameters to fine tune the text layout and placement.
%

We demonstrate the efficacy of our automated pipeline by presenting a wide variety of auto-generated example videos (\Cref{fig:teaser} and \Cref{fig:results})\footnote{Our results can be viewed at \resultsSite{}}.
We further evaluate our design guidelines and pipeline through a user study in which 57 participants rated four variations of a lyric video.
The results show that lyric videos generated by our pipeline are significantly better at maintaining text readability and unifying viewer attention.
In summary, our work makes the following contributions:
\begin{enumerate}
    \item A set of design guidelines for making lyric videos that ensure text readability and unify the viewer's focus of attention.
    \item A fully automated pipeline that instantiates these design guidelines to produce lyric videos from input music videos, which can be any video with a song as the background music.
\end{enumerate}

\begin{figure*}[t]
    \centering
    \includegraphics[width=\linewidth]{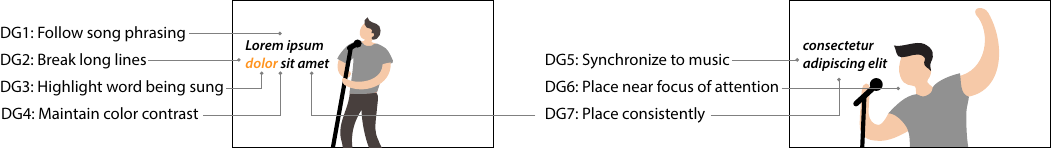}
    \caption{\emph{Illustration of the design guidelines for text readability and unified attention.}
        A lyric phrase should consist of words closely sung together (\dgPhrase), and a long text line should be broken into shorter ones consistent in length (\dgLinebreak).
        The word currently being sung should be highlighted (\dgHighlight).
        Text should be placed in areas with sufficient color contrast (\dgContrast) and near the viewer's focus of attention (\dgFocus).
        Sequential phrases should be placed in similar places (\dgSpatial).
        All lyric phrases should be synchronized to the song (\dgTiming).
    }
    \label{fig:dg}
\end{figure*}

\section{Related Work}
\label{sec:prior}

\subsection{Lyric Phrase Content and Layout}
We use the term \textit{lyric phrase} to refer to a group of lyric text that appears in the video between specific in and out times.
No work has directly studied lyric text readability in videos, so we look to prior work~\cite{Alvarez:2014:TCA, Perego:2008:WWW, Zdenek:2015:RSC, Williams:2009:BBC, DCMP:2023:GBP, Netflix:2023:ETT} on the readability of video subtitles (transcriptions of spoken words that appear at the bottom of video frame).
These works suggest that long lines of text can be hard to read, so a body of text should be segmented into phrases with appropriate line breaks for readability. 
%
Popular tools like Adobe Premiere Pro~\cite{AdobePremiere} and YouTube Studio~\cite{YouTubeStudio} use simple rule-based approaches to split a body of transcriptions into subtitle phrases based on either punctuation, character length, or temporal duration.
%

Unlike regular speeches, songs are composed of vocal phrases that group a series of lyric words together. 
Given this structure, our approach automatically organizes a song's lyrics into lyric phrases based on the temporal proximity of sequential words. 
Once a phrase has been determined, text segmentation considers where line breaks should be inserted to divide the text into more lines. 
Some studies suggest that breaking at linguistic units, such as clauses and sentences, results in better user preference~\cite{Perego:2008:WWW, Gerber:2018:LBS}.
This is not applicable to our work as song lyrics often ``do not follow formal standards of written text composition and lack punctuation''~\cite{Tegge:2020:IDT}. 
We instead break longer text into two or more lines where each line has consistent length to minimize the amount of eye movements~\cite{Perego:2008:WWW}, as excessive eye movements can distract the viewer and result in eyestrain~\cite{Hu:2015:SFV}.
Doing so also conforms to the graphic design principle of leaving no ``runt'' in new lines~\cite{Bennett:2002:MBR}. 

\subsection{Text Placement Near Focus of Attention}
Because the human eye can only read text within a small vision span~\cite{Rayner:1975:PSP}, a set of work places text near but not occluding objects that are under the viewer's focus of attention.
In subtitling, dynamic subtitles~\cite{Brown:2015:DSU} refer to subtitles placed near the speakers or other salient regions.
Hu et al.~\cite{Hu:2015:SFV} detect where the current speaker is and puts the subtitles near there.
A View on the Viewer~\cite{Kurzhals:2020:VVG} adjusts the subtitle locations live based on the viewer's eye gaze.
Kurzhals et al.~\cite{Kurzhals:2017:CAE} show that dynamic subtitles reduce the amount of eye movements and help keep the viewer's attention closer to the image content. 
Brown et al.~\cite{Brown:2015:DSU} similarly find that dynamic subtitles allowed the viewers to miss less of the video content and pick up more non-verbal cues. 
In other related domains, SmartShots~\cite{Tang:2022:SS} positions data visualizations near objects they are referencing.
SmartOverlays~\cite{Hegde:2020:SOV} places text labels near salient regions of objects detected by computer vision models to help user identify them.
%


Following this body of work, we devise a design guideline on placing lyric phrases near but not occluding objects under the focus of attention and use a combination of object detection and segmentation models in our pipeline to instantiate this.

\subsection{Tools for Adding Text to Videos}
Prior work has built tools to assist with the workflow of adding text to videos. 
Wang et al.~\cite{Wang:2017:VVS} propose an automated framework that visualizes nonverbal sounds in video with onomatopoeias (e.g. ``vroom'' for engine sounds).
The size and opacity of the added sound words are animated by the sound volume.
However, visualizing single words means that the system does not need to choose what words should be displayed together and their layout. 
Moreover, the output video is given as-is and cannot be further edited. 

EnACT~\cite{Vy:2008:EST} is a manual tool for adding animated captions to videos.
The user can annotate words in the input transcription with emotions. 
EnACT then overlays the transcription as captions on the input video and applies predefined animations to the annotated texts. 
This tool can be used to create lyric videos like ours, but the creation process remains manual, such as timing each word to the song and placing the text in the video frame.

TextAlive~\cite{Kato:2015:TLI} is a design tool for making kinetic typography videos in which lyrics are animated in synchrony with the song.
Similar to our work, it automatically aligns every word in the lyrics to the song and assigns default animations to them. 
The user can edit the animation further manually. 
However, the focus of this tools is on kinetic typography videos in which text takes the center role and does not need to accompany any underlying video content, unlike lyric videos where the coordination between text and video imagery is crucial for readability and focus of attention.

\section{Design Guidelines}
\label{sec:design}
\camready{
We follow the methodology by Agrawala et al.~\cite{Agrawala:2011:DPV} to identify lyric video design guidelines (\dg{}) from instructions, examples, and prior work (\Cref{fig:dg}). 
We first analyzed 15 text tutorials, 5 video tutorials, and 3 subtitle guidelines~\cite{Zdenek:2015:RSC, Williams:2009:BBC, DCMP:2023:GBP} to form draft guidelines if similar content appeared repeatedly.
While a few of these guidelines were concrete (\dgContrast{} and \dgTiming{}), others remained vague at this stage. For example, \dgPhrase{} and \dgLinebreak{} were about ``having some words in one or two lines.'' \dgFocus{} was ``don't obstruct video content.''

To formalize the design guidelines, we analyzed the top 100 most viewed lyric videos to find patterns (full video list in supplemental materials).
For example, \dgSpatial{} is a recurring observation on text placement. We also observed common highlighting animations to use as \dgHighlight{}'s default options.
Prior work also contributed to the guideline definitions. Reducing eye movements~\cite{Hu:2015:SFV,Kurzhals:2017:CAE,Perego:2008:WWW} helped define \dgLinebreak{}, \dgFocus{}, and \dgSpatial{}.
Graphic design layout principle~\cite{Bennett:2002:MBR} contributed to \dgLinebreak{}, and sheet music composition to \dgPhrase{}.
}

%

We present the set of design guidelines below. The two overarching goals of our guidelines are (1) ensuring the readability of the added lyric text and (2) maintaining a unified focus of attention.


\subsection{Text Readability}
Good text readability can be achieved by properly composing words into lyric phrases with line breaks.
Moreover, presenting the text with sufficient contrast and animated highlighting can help guide the viewer's eyes to quickly locate the right words.

\begin{description}[style=unboxed,leftmargin=0cm]
    \item[\dgPhrase{}: A lyric phrase should consist of words sung closely together.]
    In sheet music, a phrase mark (slur) spans over a set of notes to indicate that they should be sung together as a phrase.
    These musical phrases are the building blocks of a song.
    Therefore, a lyric phrase should respect such phrasing by incorporating words in the same musical phrase and display them in the video as a unit.\\


    \item[\dgLinebreak{}: Long text in a lyric phrase should be broken into two or more lines with consistent length.]

    Some lyric phrases might contain many words, such as in a fast-paced song.
    A long line of text is slower and harder to read because it requires excessive eye movement; it can also be distracting when added into the video, as suggested by prior research~\cite{Alvarez:2014:TCA}.
    %
    %
    Because of this, a long line should be broken in two or more lines with consistent length.
    Doing so minimizes the amount of eye movement when reading from line to line~\cite{Perego:2008:WWW} and also conforms to the graphic design principle of leaving no ``runt'', a single or few words at the end of a paragraph~\cite{Bennett:2002:MBR}.\\



    \item[\dgHighlight{}: Lyric text should be highlighted as it is sung.]
    %
    %
    As human eyes are sensitive to changes in state such as color and motion, applying animated highlighting to the word currently being sung can guide the viewer's eyes to more quickly see and read the right word.\\
    %
    %





    \item[\dgContrast{}: Lyric text should have sufficient contrast against its background.]
    A sufficient color contrast between the text and its background is important for readability.
    Alternatively, contrast can also be achieved by styling the text, such as adding outlines, drop shadow, or semi-transparent background boxes.
\end{description}

\subsection{Viewer Attention}
As the added text requires the viewer's attention to read and comprehend, it is important to coordinate the text with the song audio and video imagery to not split the focus of attention.

\begin{description}[style=unboxed,leftmargin=0cm]
    \item[\dgTiming{}: Lyric text should be synchronized to the song.]
    If the lyrics and song are misaligned in timing, the viewer may be distracted by processing similar information more than once.
    This can cause them to miss important details such as visual cues in the video, negatively affecting their understanding of the content.\\

    \item[\dgFocus{}: Lyric text should be placed near but not occluding the focus of attention.]
    Prior work has shown that extensive eye movements distract the viewer from understanding details in the video~\cite{Kurzhals:2017:CAE} and contribute to eye strain~\cite{Hu:2015:SFV}.
    Thus, text should be placed near the focus of attention to minimize the viewer's eye movement.
    However, it should also not occlude objects in the focused region as their actions can be crucial to the understanding of the video content. For example, a singer might also be dancing when singing, so the text should avoid occluding any part of the singer's body. Speech bubbles in comics and manga are examples of this idea applied in other media formats.\\



    \item[\dgSpatial{}: Sequential lyric phrases should be placed near each other.]
    Because a lyric phrase is displayed in the video for a limited amount of time, sequential lyric phrases should be placed near each other so that they can be seen without searching. Doing so also minimizes the amount of eye movement~\cite{Hu:2015:SFV}.

    \begin{figure*}[th]
    \centering
    \includegraphics[width=\linewidth]{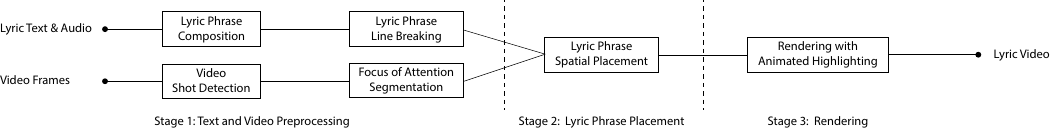}
    \caption{
        Our pipeline consists of three stages.
        Given a music video as input, stage 1
        aligns the lyrics to the song and groups it into phrases with line breaking.
        This stage also produces segmentation masks of the objects under the focus of attention.
        Stage 2 computes the spatial placement of the text via an optimization approach considering terms related to color contrast, attention masks, and the position of the previous phrase.
        Stage 3 renders the final lyric video with animated highlighting.
    }
    \label{fig:stages}
\end{figure*}

\end{description}


\begin{figure}[htb]
    \centering
    \includegraphics[width=\linewidth]{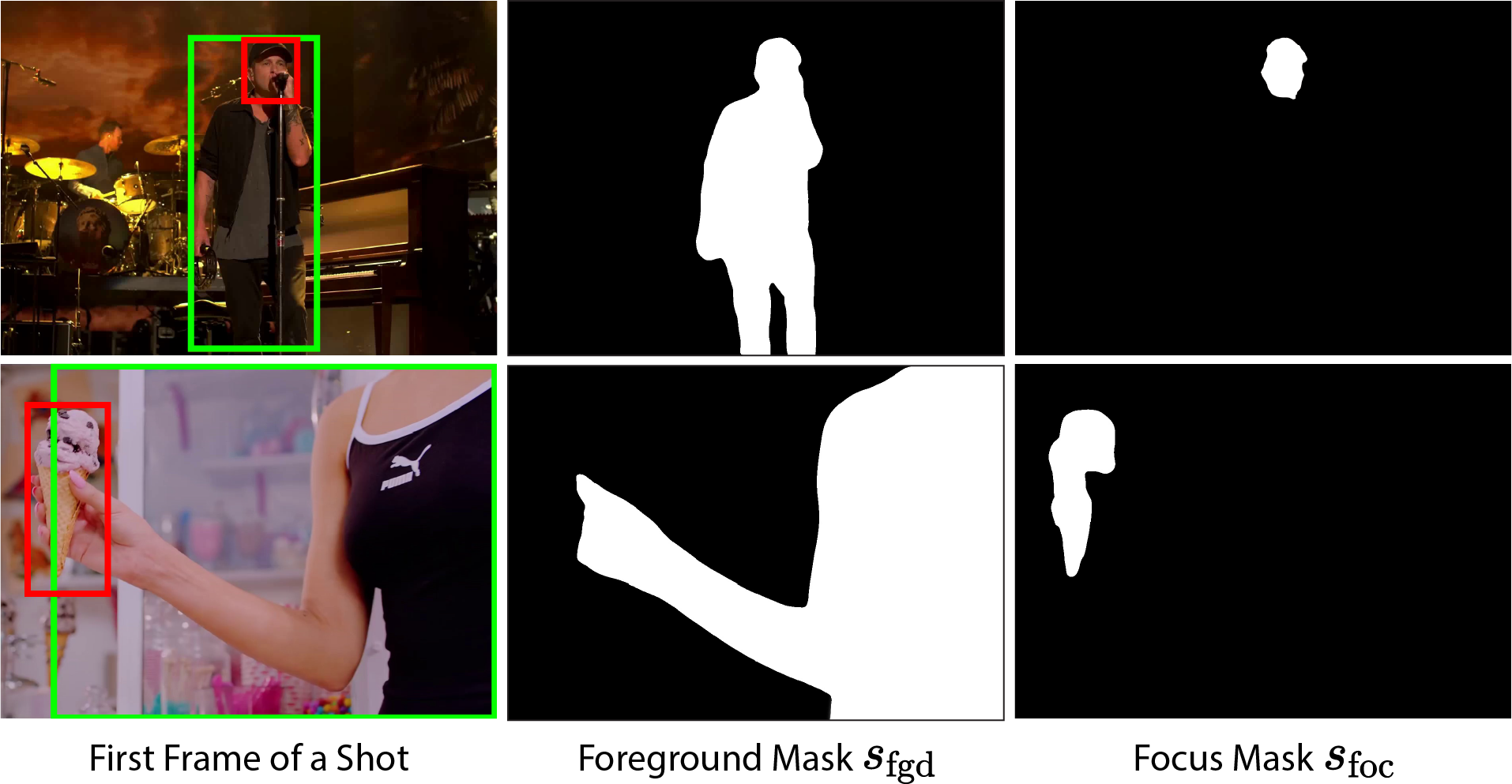}
    \caption{For every frame, we generate two segmentation masks.
    In the absence of a person's body or face, we look for objects referenced by the lyrics (Row 2).
    We place a lyric phrase near the focus mask while not occluding the foreground mask.
    \camready{%
    \emph{%
        Video source: 
        I Ain't Worried by OneRepublic~\cite{OneRepublic:2022:IAW},
        Ice cream by BLACKPINK \& Selena Gomez~\cite{Blackpink:2021:IC}.
    }
    }
    }
    \label{fig:saliency}
\end{figure}

\begin{figure*}[h]
    \includegraphics[width=\textwidth]{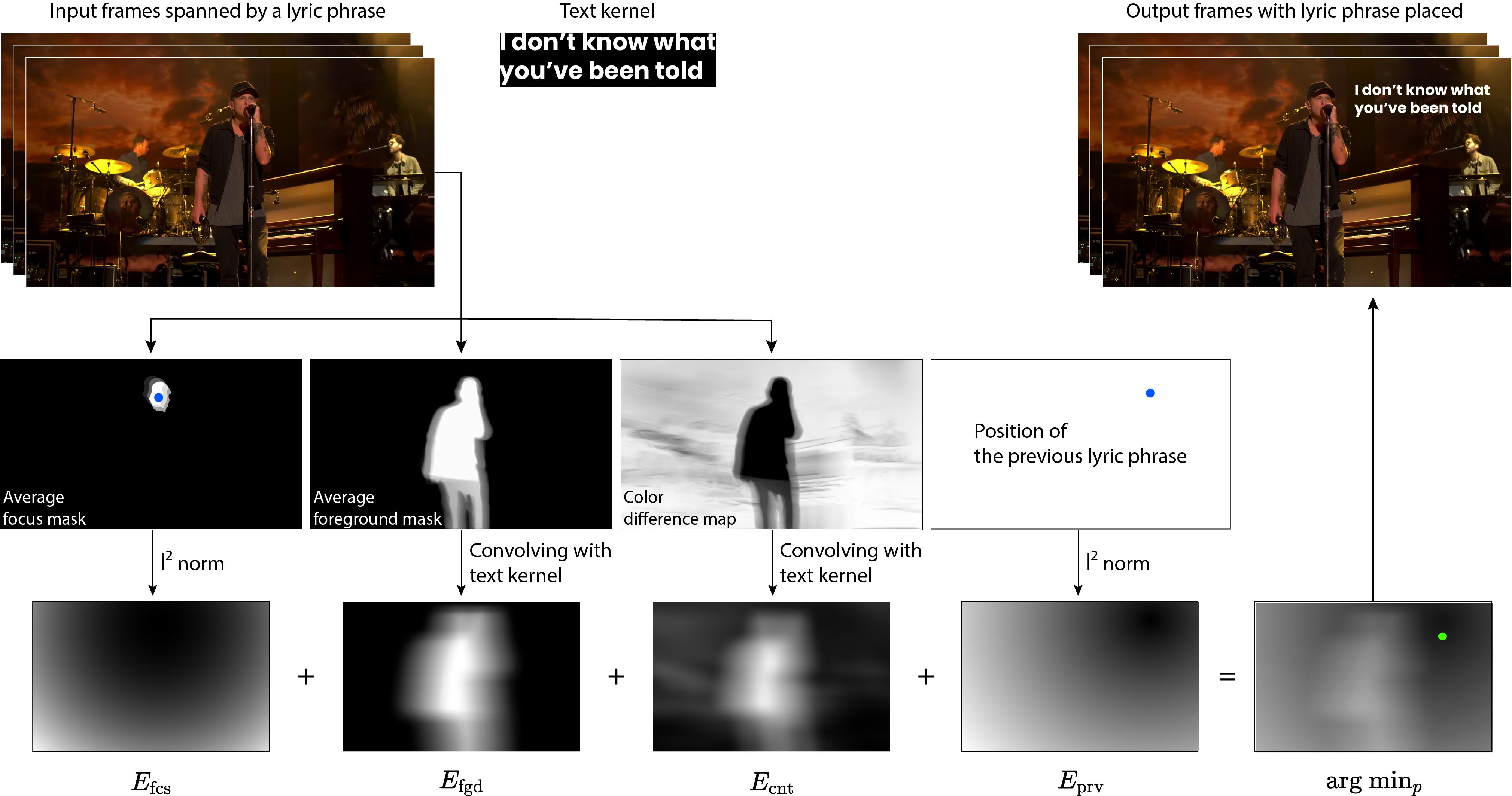}
    \caption{\emph{Visualization of the lyric phrase placement algorithm.}
        Given a lyric phrase and its associated set of video frames, the placement algorithm minimizes a linear combination of energy functions related to the design guidelines to find the optimal position for that lyric phrase.
        Note that darker pixels correspond to lower scores, and the cost maps associated with the energy functions are smaller than the input frame size because we do not allow the text to be placed partially or fully out of the screen.
        \camready{%
            \emph{%
            Video source: 
            I Ain't Worried by OneRepublic~\cite{OneRepublic:2022:IAW}.}
        }
    }
    \label{fig:energy_maps}
\end{figure*}

\section{Automated Pipeline}
\label{sec:method}


With the set of design guidelines identified, we have developed a fully automated pipeline that instantiates these guidelines to convert a music video into a lyric video.
\camready{The input is a video with a song as its background music,} such as music videos released by the artists, recordings of live performances, and fan-made lip sync videos and covers.
By default, the typeface of the text is Poppins with regular weight, and it is white and 40 pixels in size.
The user can adjust the style based on their preferences.

As shown in \Cref*{fig:stages}, our pipeline has three stages: text and video contents are preprocessed in stage 1, the spatial placement of the text is computed via an optimization approach in stage 2, and the final lyric video is rendered with animations in stage 3.
%
In stage~1, text and video preprocessing ,we group the song's lyrics into phrases with line breaking (\dgPhrase\ and \dgLinebreak) and obtain the in and out timing of each word (\dgTiming).
We also compute segmentation masks for objects under the focus of attention for every frame of the input video (\dgFocus).
In stage 2, text placement, we generate the spatial position of each lyric phrase by solving an optimization problem that minimizes energy functions related to saliency (\dgFocus), color contrast (\dgContrast), and spatial consistency of positions of sequential phrases (\dgSpatial).
Finally, in stage 3, rendering, we render the output lyric video with animated highlighting (\dgHighlight) based on the spatial and temporal information computed in earlier stages.

\subsection{Stage 1: Text and Video Preprocessing}
In this stage, lyric phrases are generated by grouping words that are close in time with line breaks added to long text.
On the video side, we find and mask out the objects under the attention in every frame.
We use the resulting segmentation masks to determine the optimal positions of texts in stage 2.

\subsubsection{Generating lyric phrases}
We first fetch the song lyrics of the input video from Musixmatch~\cite{Musixmatch}.
We then use AutoLyrixAlign~\cite{Gupta:2020:ALA} to obtain word-level temporal alignment, a pair of time in seconds that specify the in and out times, for each lyric word.
In accordance with \textbf{\dgPhrase}, we group lyric words into individual phrases based on their temporal proximity to each other.
More specifically, sequential words are assigned to the same lyric phrase if the difference between the out time of the previous word and the in time of the next word is within a set threshold.
By default, the value of this threshold is set to the length of a beat, detected via librosa~\cite{McFee:2015:LAM}. Given that there might be inaccuracies in the timing of the words, the user can adjust this threshold.
We set the timing of the lyric phrase to start at the in time of its first word and end at the out time of its last word.

\subsubsection{Adding line breaks to lyric phrases}
To satisfy \textbf{\dgLinebreak}, we add line breaks to lyric phrases with long lines of text.
We count the number of characters of every lyric phrase, set the median length as the threshold, and break phrases longer than this threshold into lines that are close in length.
Similar to the threshold for grouping phrases, the user can adjust the threshold for line breaking.

\subsubsection{Video Shot Detection}
As a video is composed of shots and visual contents in a single shot are more consistent, it is easier to extract segmentation masks for objects under attention from each shot separately instead of the entire video.
Therefore, we split the input video into successive shots via the approach of Rao et al.~\cite{Rao:2020:LGA}.

\subsubsection{Focus of Attention Segmentation}
\label{subsubsec:focus}
As human attention is driven in part by ``task at hand and current goals'' (top-down attention~\cite{Katsuki:2013:BUT}) and drawn to sounding objects~\cite{Chen:2023:ACS}, we thus look for human figures and objects referenced by the lyrics as objects under the focus of attention.
We refer to these objects as foreground objects.

To identify the human figures in the video, we first use an object instance detection model~\cite{feng2021tood} to obtain bounding boxes of people in the first frame of every shot.
For every person instance detected, we run a face detection model and extract their facial features represented as a 1D vector~\cite{huang2018unifying}.
If the cosine similarity of two facial features in two shots is lower than $0.1$ (determined empirically), we assign the same person label to these two instances.
We then count the number of appearances of each distinct person.

In each shot's first frame, the person with the highest appearance frequency becomes the foreground object, and we input the bounding boxes of their body and face to a video object segmentation model~\cite{Paul:2022:RTS} to obtain segmentation masks for their body and face for every frame.
%
%
%
In the absence of people or their faces (\Cref{fig:saliency}), we look for any object (noun) referenced by the lyric phrases using another object instance detection model pretrained on Objects365~\cite{shao2019objects365}.
If no suitable person nor referenced object is detected (e.g. the shot is a b-reel for filler purposes), the output masks are completely black.
%
%
As shown in \Cref{fig:saliency}, we refer to the body segmentation mask as the foreground mask $\foregroundMask$ and the face mask as the focus mask $\focusMask$.

\subsection{Stage 2: Lyric Phrase Placement}
As shown in \Cref{fig:energy_maps}, in this stage, we find the optimal position $p_{\text{min}}$ of a lyric phrase by minimizing a linear combination of energy functions implemented according to the design guidelines.

To compute values for the terms in our total energy function, we first collect the set of frames $\spannedFrames$ spanned by the in and out times of a lyric phrase.
For the frames in $\spannedFrames$, we compute the pixel-wise average of their focus mask $\focusMask$ and foreground mask $\foregroundMask$ (\Cref{fig:saliency}) to obtain an average focus mask $\avgFocusMask$ and an average foreground mask $\avgForegroundMask$ (Row 2 of \Cref{fig:energy_maps}).
Furthermore, we obtain the background $b$ of a frame $f$ by inverting the foreground mask and multiplying by $f$ so that $b = f \cdot (1 - \foregroundMask)$.
We then compute the pixel-wise average of all backgrounds of frames in $\spannedFrames$ to obtain an average background image $\bar{b}$.
We compute averages because we optimize for a fixed text position over the time span of a lyric phrase.
%
Finally, we rasterize the text of a lyric phrase into a greyscale image $\textkernel$, stored as a 2D array of floats ranging from 0 to 1.
We use $\textkernel$ as a convolution kernel in the energy terms.
We describe specific energy terms below, with $p$ denoting a candidate pixel coordinate for placing the lyric phrase (upper left corner of the text rectangle).

\paragraph{Placement near focus of attention}
The first two energy functions in our optimization, $\energyVisualCenter$ and $\energyOverlap$, correspond to \dgFocus.
$\energyVisualCenter$ puts the lyric phrase position $p$ close to the visual center of mass $p_{\text{center}}$ of $\avgFocusMask$,
which is the average position of the pixel intensities in $\avgFocusMask$:
$$\energyVisualCenter(p) = ||p_{\text{center}} - p||_2$$

$\energyOverlap$ ensures that the lyric text minimally occludes the white regions in the average foreground mask $\avgForegroundMask$.
To achieve this, we convolve the text kernel $\textkernel$ with $\avgForegroundMask$ to produce a foreground overlap cost map $o = \avgForegroundMask \otimes \textkernel$.
Since entries in $\textkernel$ have values from 0 to 1, $o(p)$ returns the weighted sum of the pixels in $\avgForegroundMask$ that overlaps with the text if it is placed at $p$.
Therefore, the value of $o(p)$ is lower when there is less overlap (\Cref{fig:energy_maps}).
Our energy function $\energyOverlap$ is then computed as:
$$\energyOverlap = o(p)$$

\paragraph{Placement with high color contrast}
$\energyContrast$ places the text against background regions with high color contrast (\dgContrast).
Similar to $\energyOverlap$, we compute a background color difference cost map $c(p)$ that returns the color difference value between the text and its background if it is placed at position $p$.
To obtain $c$, we first compute a background color difference image $\bar{b}_{\text{diff}}$ by subtracting the text color from the average background $\bar{b}$ (Row 2 of \Cref{fig:energy_maps}).
The difference between two colors is defined as the Euclidean distance between two RGB colors.
We then convolve the color difference map with the inverted text kernel $1 - \textkernel$.
We invert the kernel so that the color differences of pixels surrounding the text, instead of underneath the text, are taken into consideration.
The resulting $c' = \bar{b}_{\text{diff}} \otimes (1 - \textkernel)$ has entries with values equal to the sum of the color differences surrounding the text.
We then invert $c'$ by subtracting its maximum value from it to obtain $c$ so that lower value means higher color contrast.
Our energy function $\energyContrast$ is thus: $$\energyContrast = c(p)$$

\paragraph{Placement near previous lyric phrase}
The energy function $\energyPrev$ is designed to place each lyric phrase near the previous one (\dgSpatial).
Given the position of the previous phrase $p_{\text{prev}}$, $\energyPrev$ is defined as:
$$\energyPrev = ||p_{\text{prev}} - p||_2$$
Note that, for the very first lyric phrase and the first lyric phrase in a shot, we set $\energyPrev = 0$.
We do not let the previous phrase influence the positioning of the current one because the composition of video imagery often changes significantly from shot to shot.

Together, we combine the individual energy functions described above to define the following optimization function:
\begin{align*}
    p_{\text{min}} = \text{arg} \ \text{min}_{p} \left(
        w_{\text{fcs}} \energyVisualCenter +
        w_{\text{fgd}} \energyOverlap +
        w_{\text{cnt}} \energyContrast +
        w_{\text{prv}} \energyPrev \right)
\end{align*}
where each weight $w$ is adjustable for fine-tuning the positioning.

\subsection{Stage 3: Rendering}

At the end of stage 2, every lyric phrase has a coordinate $p$ for its placement in the video and in and out times for the whole phrase and each individual word.
We write a script in Adobe ExtendScript to automatically parse and load the lyric phrase data into After Effects and overlay them with animated highlighting (\textbf{\dgHighlight}) onto the original music video.
To further support \textbf{\dgFocus}, we use the foreground segmentation masks to overlay the foreground objects on top of the text.
The user can choose not to apply these features.

We provide a set of commonly used highlighting animations (\textbf{\dgHighlight}) designed based on existing lyric videos.
All text in a phrase can be animated in and out together, either through fading, sliding up or down, or a combination of both.
We add extra padding time (0.2 seconds, adjustable) to the in and out times of a phrase for the animations to take place.
Doing so also gives the viewer extra time to read the last few words in a phrase if they are sung shortly.
Individual words can be highlighted via fading in and out, sliding up and down, changing to an accent color, or a combination of them.
The default highlighting is that a phrase fades in and out with individual words sliding up when sung.
We apply these animations by automatically inserting keyframes at appropriate times via ExtendScript.
Rendering the video in After Effects additionally allows the user to edit it further by taking advantage of the comprehensive set of tools that After Effects provides.

\camready{
\subsection{Example Usage Scenario}
To obtain a lyric video, the user first inputs a music video into the pipeline for a result.
They can then adjust the default settings by editing values in a JSON file.
For example, the user can adjust the text font, color, and animation, or increase weights such as $w_{\text{fcs}}$ and $w_{\text{fgd}}$ to strongly draw the text to the focus of attention (\dgFocus).
They can also fix pipeline errors as described in Section~\ref{sec:cases}.
After the desired adjustments are made, they can rerun the pipeline for a new result. They can repeat this process multiple times, or make one-off or more detailed adjustments in After Effects.
}
\begin{figure*}[t]
    \centering
    \includegraphics[width=\linewidth]{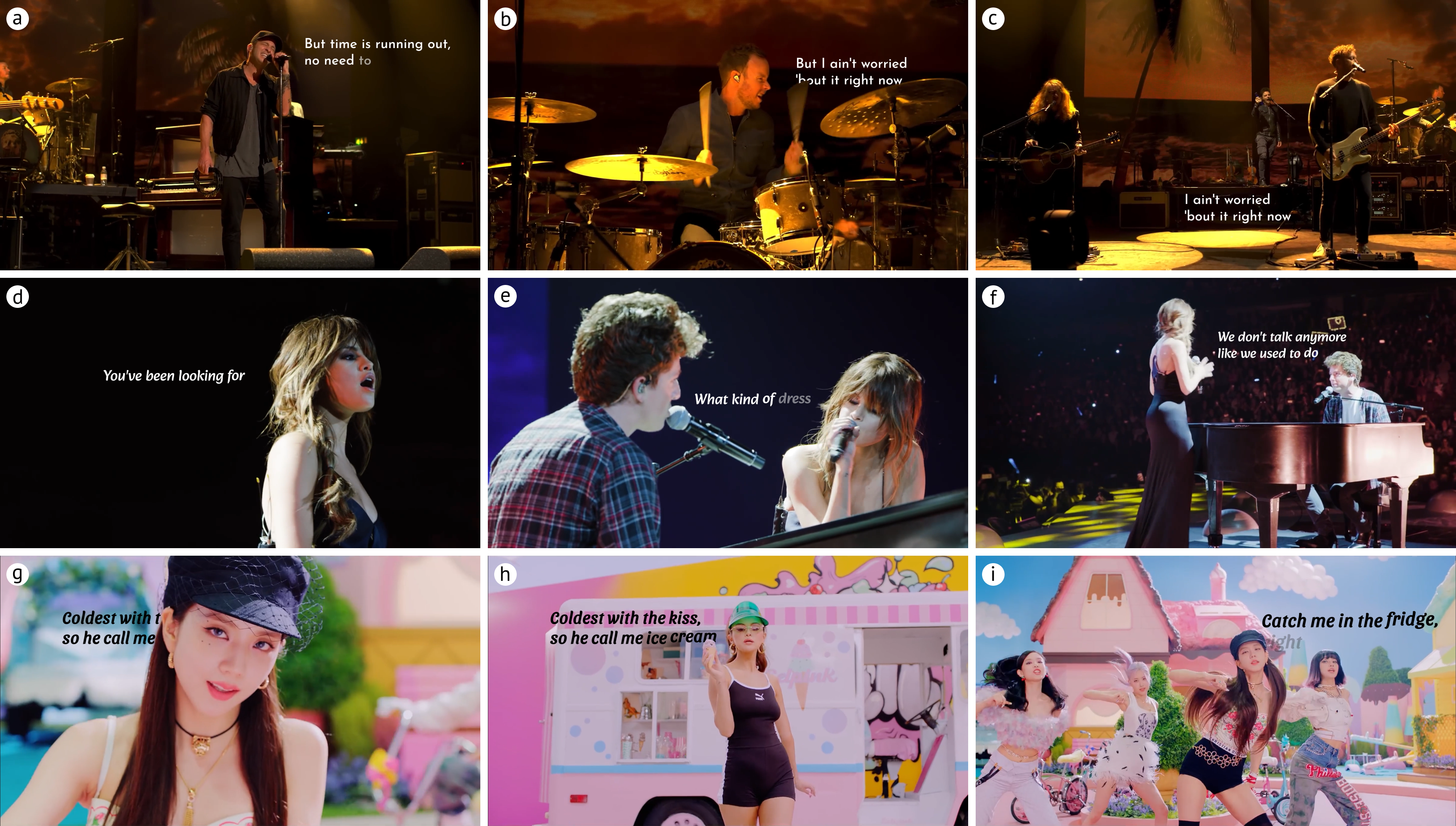}
    \caption{We present 3 lyric videos automatically generated by our pipeline from inputs that are challenging to add text to.
        In the first video (a-c), the camera constantly switch among the musicians while also zooming in and out.
        In the second video (d-f), two singers are present and the main female singer constantly walks around the stage.
        The third video (g-i) features many quick shot changes with significant differences in composition, as well as many musicians.
        In these cases, our pipeline is able to generate results that adhere closely to our design guidelines.
        \camready{%
        \emph{%
            Video sources (from top to bottom):
            I Ain't Worried by OneRepublic~\cite{OneRepublic:2022:IAW},
            We Don't Talk Anymore by Charlie Puth \& Selena Gomez~\cite{Puth:2016:WDTA},
            Ice Cream by BLACKPINK \& Selena Gomez~\cite{Blackpink:2021:IC}.}}
    }
    \label{fig:results}
\end{figure*}

\section{Results}
\label{sec:result}

To demonstrate the efficacy of our automated pipeline, we have generated 15 lyric videos from music videos on YouTube. \camready{Please find the resulting videos at \resultsSite{}.}

Figure~\ref{fig:teaser} and Figure~\ref{fig:results} feature 9 of the resulting videos.
We select a variety of input videos, including official music videos, live performances (\Cref{fig:teaser}c, \Cref{fig:results}a-f), and animated features (\Cref{fig:teaser}d).
The songs are from diverse genres with tempos ranging from slow to fast.
The arrangement of the musicians varies from one singer, a singer with a band, to multiple singers (\Cref{fig:results}).
We also use videos in both horizontal and vertical aspect ratios (\Cref{fig:teaser}c and~f).
All the results presented are generated by our fully automated pipeline without any manual edits in After Effects (some input parameters, like text style and animation, are adjusted for certain examples) .

In all these examples, our pipeline consistently finds and places text next to the viewer's focus of attention ($\energyVisualCenter$ for \dgFocus), such as the ice cream cone held by the singer in \Cref{fig:results}h.
In the case of \Cref{fig:teaser}a where multiple faces are present, our pipeline also correctly identifies the main singer's face via appearance frequency counting.
Our pipeline is also able to identify spare regions, an area with minimal foreground actions ($\energyOverlap$ for \dgFocus) and good color contrast ($\energyContrast$ for \dgContrast) in the video.
In \Cref{fig:teaser}b, our pipeline places the text in the dark-colored window right next to the singer.
In \Cref{fig:teaser}d, as the singer swings her arm from the lower left to top right, the pipeline finds the spare region in between her face and her arms for the text.
Similarly, the text sits at a nice dark area near the center of the video frame \Cref{fig:results}c.

In addition to horizontal aspect ratio videos, our pipeline can also add text to vertical videos that are popular on social media platforms like TikTok and Instagram.
\Cref{fig:teaser}c and f are two such examples. By breaking a lyric phrase into multiple lines, our pipeline is able to fit the text into the narrow width of the video and places it near the singer with good contrast.

\Cref{fig:results} presents three examples whose input music videos are challenging to process.
In the first video of a live performance (\Cref{fig:results}a-c), the singer is accompanied by a band of musicians, and the video camera switches back and forth to feature different people while also zooming in and out.
Our pipeline can consistently identify the other musicians in the absence of the singer (\dgFocus) and avoid bright-colored regions (\dgContrast) for placing white text (\Cref{fig:results}b-c).
In the second video (\Cref{fig:results}d-f),
the main female singer walks around the male singer playing the piano, and the camera occasionally switches to showing the concert audience.
Our pipeline places the text to consistently follow the female singer and finds space for the lyric text in between the two singers when they are close together (\Cref{fig:results}e-f).

The third video (\Cref{fig:results}g-i) is even more challenging than the previous two in that it is a fast-tempo song featuring 5 singers with dynamic dance movements.
Shot changes occur very frequently, as one lyric phrase often spans three or more shots.
\Cref{fig:results}g-h shows an example of a lyric phrase spanning multiple shots. In \Cref{fig:results}g, the lyric phrase is briefly, partially blocked by the foreground object.
Such occlusion ensures that the text does not distract the viewer from the focus of attention, while the impact on readability is minimal since the word currently being sung (``me'') is still visible.
After just a few frames, the shot quickly changes to the one shown in \Cref{fig:results}h.
Even though the singer's position changes and the camera zooms out, the text remains near the focus of attention.

Overall, given a wide variety of input videos, our pipeline is able to produce lyric videos that closely follow the design guidelines and makes the text look like it is an integral part of the video.
\begin{figure*}[t]
    \includegraphics[width=\textwidth]{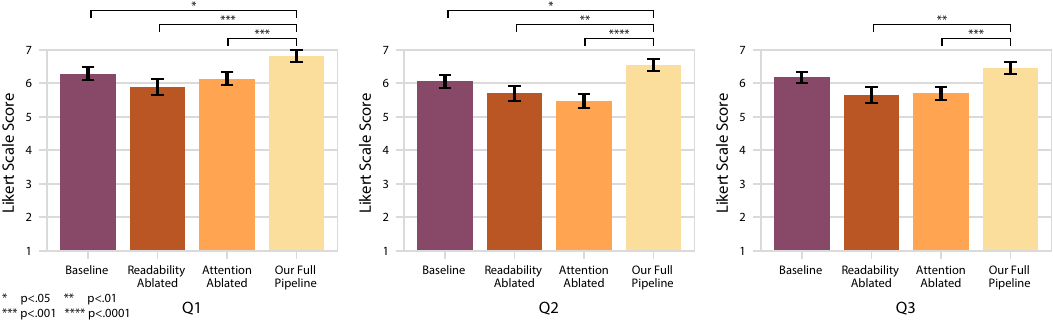}
    \vspace{-15pt}
    \caption{
        \emph{
            Aggregated Likert scale ratings of questions on text readability (Q1), unified attention (Q2), and overall experience (Q3).
        }
        Horizontal brackets indicate pairwise significant difference, and the error bars show standard error.
        Overall, the results demonstrate that the lyric videos produced by our pipeline are significantly better at achieving the goals of ensuring lyrics readability and unifying the viewer's focus of attention than the other versions.
    }
    \label{fig:likert}
\end{figure*}

\section{Evaluation}
\label{sec:evaluation}
We conducted a user evaluation to investigate the following research question: How well does the lyric video produced by our automated pipeline achieve the two overarching goals of our proposed design guidelines, ensuring text readability and maintaining unified focus of attention?

\subsection{Materials}
To examine the research question, for a given input music video, we generated 4 conditions: a \emph{Baseline} condition in the style of video subtitles, a \emph{Full} condition generated by our automated pipeline, and two conditions, \emph{Readability Ablated} and \emph{Attention Ablated}, generated by pipelines in which we ablated certain components based on relevant design guidelines.

The Baseline condition is made to look like video subtitles. Each lyric phrase corresponds to one line in the original lyrics file retrieved from Musixmatch with no line breaks and always sits at the bottom center of the video.
The Readability Ablated condition does not incorporate design guidelines related to text readability (\dgPhrase{}-\dgContrast{}); in this version, a lyric phrase is composed in the same way as the Baseline condition.
We do not add animated highlighting and do not include the color contrast energy function (i.e., we set $w_{\text{cnt}} = 0$) in spatial placement.
The Attention Ablated condition does not consider design guidelines on unifying attention (\dgFocus{}-\dgSpatial{}).
We still sync the lyric phrases to the song, but we do not place them near the focus of attention (i.e., $w_{\text{fcs}} = w_{\text{fgd}} = 0$) and sequential lyric phrases are not placed near each other (i.e., $w_{\text{prv}} = 0$).

\camready{We chose 10 of the input music videos from Section~\ref{sec:result} for the evaluation (full video list in Section~\ref{sec:list})}
and created a total of 40 videos.
The generated videos range from 30 to 60 seconds in length (only sections of the original videos are used to control the total duration of the survey).



\subsection{Procedure}
We distributed our online survey via multiple listservs for a wide range of participants from the US.
Each survey presented a participant with the 4 conditions of one video randomly chosen from the 10 videos in our evaluation set.
The viewing sequence of the 4 conditions was randomized to mitigate the learning effect.
After viewing each video, we asked the participant to rate the following statements on a 7-point 
Likert scale and elaborate on their choices in a free response form.
\begin{description}
    \item[Q1:] The text of the lyrics in the video is easy to read.
    \item[Q2:] I can both easily read the text and watch the video imagery.
    \item[Q3:] The overall viewing experience is good.
\end{description}
The scale ranged from Strongly Disagree (1) to Strongly Agree (7) for all questions.
At the end of the survey, we also collected the participant's age and gender.
The participants did not receive any monetary compensation for completing the survey.

\subsection{Evaluation Results}
We received 57 valid responses to the online survey after removing duplicated answers.
The participants (27 female, 28 male, 1 non-binary, 1 prefer not to say) range from 19 to 32 years old ($\bar{x} = 25.35, \text{SD} = 2.03$).
Each of the 10 videos was shown to at least 5 participants.  
We aggregate ratings on the same Likert scale question for the same condition of a video and show the results in \Cref{fig:likert}. 

For each of the three questions (Q1-3), we first conducted a Friedman's Test which shows that there is significant difference among the 4 conditions ($p < 0.001$ for all questions).
We then ran post-hoc pairwise two-tailed Wilcoxon signed-rank tests with Holm-Bonferroni correction to compare the Full condition against the other three.
The full Wilcoxon statistics can be found in Table~\ref{table:stats} in the appendix.
For Q1, the Full condition ($\bar{x} = 6.80$) 
is rated significantly higher than the other three: 
vs. Baseline ($\bar{x} = 6.28$): $p = 0.043$; 
vs. Readability Ablated ($\bar{x} = 5.89$): $p = 0.0004$; 
vs. Attention Ablated ($\bar{x} = 6.14$): $p = 0.0003$.
For Q2, the Full condition ($\bar{x} = 6.54$)
performances significantly better than the other three:
vs. Baseline ($\bar{x} = 6.05$): $p = 0.034$;
vs. Readability Ablated ($\bar{x} = 5.70$): $p = 0.001$;
vs. Attention Ablated ($\bar{x} = 5.47$): $p = 0.000006$.
For Q3, the Full condition ($\bar{x} = 6.45$) 
is rated significantly higher than the two ablated conditions: 
vs. Readability Ablated ($\bar{x} = 5.64$): $p = 0.003$;
vs. Attention Ablated ($\bar{x} = 5.70$): $p = 0.0006$,
and is not significantly different from the Baseline ($\bar{x} = 6.17$).

Overall, these results indicate that the Full condition produced by our pipeline is significantly better than the ablated conditions in terms of maintaining readability, unifying attention, and overall experience (\Cref{fig:likert}).
Moreover, the Full condition is also significantly better than the Baseline condition in the first two measures. 


The participant comments provide insights into the quantitative results.
For the Full version, without knowing the design guidelines, 14 participants specifically commented that the placement of the text is close to their focus of attention: ``The text is very close to where my attention of the video would be'' (P22).
This helped them to both easily read the text (``Text was easy to follow'', P19) and pay attention to the text and video together (``It was easy to saccade back and forth'', P25).
P57 mentioned that placing sequential phrases in the same shot near each other helps with readability: ``Easier to find this time as some of them start in the same place.''

In the Readability Ablated version, 11 participants found that the text is still near their focus of attention, but certain phrases are too long to be easily read and overlap with objects\liyi{(July 21, 2023) shouldn't this be under the attention-ablated group instead of readability-ablated group?} under their attention: ``the lyrics will occlude the main subject and each segment is too long'' (P38).
P29 additionally found that the lack of highlighting animation negatively affects the text readability: ``Locating the text was also harder because there's no movement.''

The main issue with the Attention Ablated condition is that lyric phrases are placed often far away from the focus of attention, making them hard to find.
11 participants mentioned this concern: 
``Can't really focus on the video because I have to spend time looking for the text '' (P46), and ``Can only focus on either one'' (P42).
On a positive note, 4 participants did find the animated highlighting, which is also applied in the Full version, helpful for readability: ``The way each word was emphasized along with the singing freed up some processing load'' (P44).

\camready{
The participants rated that the Baseline condition provides an overall good viewing experience similar to the Full condition.
Familiarity with the subtitle text contributes to the Baseline condition's high rating, as 12 participants mentioned that they knew the text would appear at a fixed location: ``This pattern is very comfortable and common'' (P57). However, some also noted that this trades creative expression for predictability (P53: ``this feels like subtitles, not like a lyric video'').
%
Despite the similar ratings, the participants used more positive words for the Full condition, such as ``enjoy'', ``easy'', and ``fun'', and instead described Baseline as ``normal'', ``standard'', and ``familiar'' (P19: ``not an outstanding viewing experience'').
}
Moreover, the Baseline condition also shares the same issue as the Attention Ablated version.
8 participants stated that their eyes need to move around a lot in order to read the text and watch the video: ``Sometimes I need to skip some lines to focus on the singer'' (P10) and ``Eyes have to move a lot to look at other places'' (P3).

Overall, significant differences in the Likert scale ratings and the participant comments demonstrate that the lyric video generated by our automated pipeline, which instantiates our design guidelines, are effective in achieving our two overall goals, ensuring text readability and unifying the viewer's focus of attention.





\section{Limitations and Future Work}
\label{sec:limitations}
\camready{We acknowledge that the 10 input videos in our user evaluation forms a relatively small corpus, and there is a lack of comparison to user-made lyric videos.}

\camready{
We discuss common failure cases of our pipeline in detail in Section~\ref{sec:cases}.
One area that future work can focus on is improving our focus of attention segmentation algorithm.
%
Future work can explore integrating deep learning-based saliency detection models that use bottom-up attention cues, such as motion and color, into our current top-down approach.
Moreover, a multi-modal approach can be considered, such as identify sounding instruments or the singer of the current phrase in case of multiple singers.
}

\camready{
Our pipeline works best when there are enough spare regions (places with minimal foreground actions and sufficient color contrast)
in the input video (more discussed in Section~\ref{sec:cases}).
}
Some stylized videos are purposefully designed this way, and
many vertical videos also do not have such regions since a human face or body can often take up the majority of the screen.
Future work might explore means to modify existing video content to make space for lyric text, such as finding and blurring the region in a frame with the least amount of visual information.

\section{Conclusion}
\label{sec:conclusion}
Lyric videos are widely produced today despite the amount of time and careful coordinations they require to make.
We identify 7 design guidelines to help creators ensure that the text in these videos are readable and the viewer's focus of attention are unified.
We further implement a fully automated pipeline that converts an input music video into a lyric video following these design guidelines.
We demonstrate the efficacy of our pipeline by generating lyric videos from music videos that vary significantly in format and imagery.
A 57-respondent user study shows that lyric videos produced by our pipeline are effective in achieving our goals of ensuring text readability and maintaining unified focus of attention.
\begin{acks}
\camready{
We would like to thank John Nelson for his inputs during the early stage of this project.
This work was supported by the Stanford Graduate Fellowship and Brown Institute for Media Innovation.
}
\end{acks}


\bibliographystyle{ACM-Reference-Format}

\bibliography{paper,misc}

\newpage 
\appendix
\normalsize

\begin{figure}[htb]
    \centering
    \includegraphics[width=\linewidth]{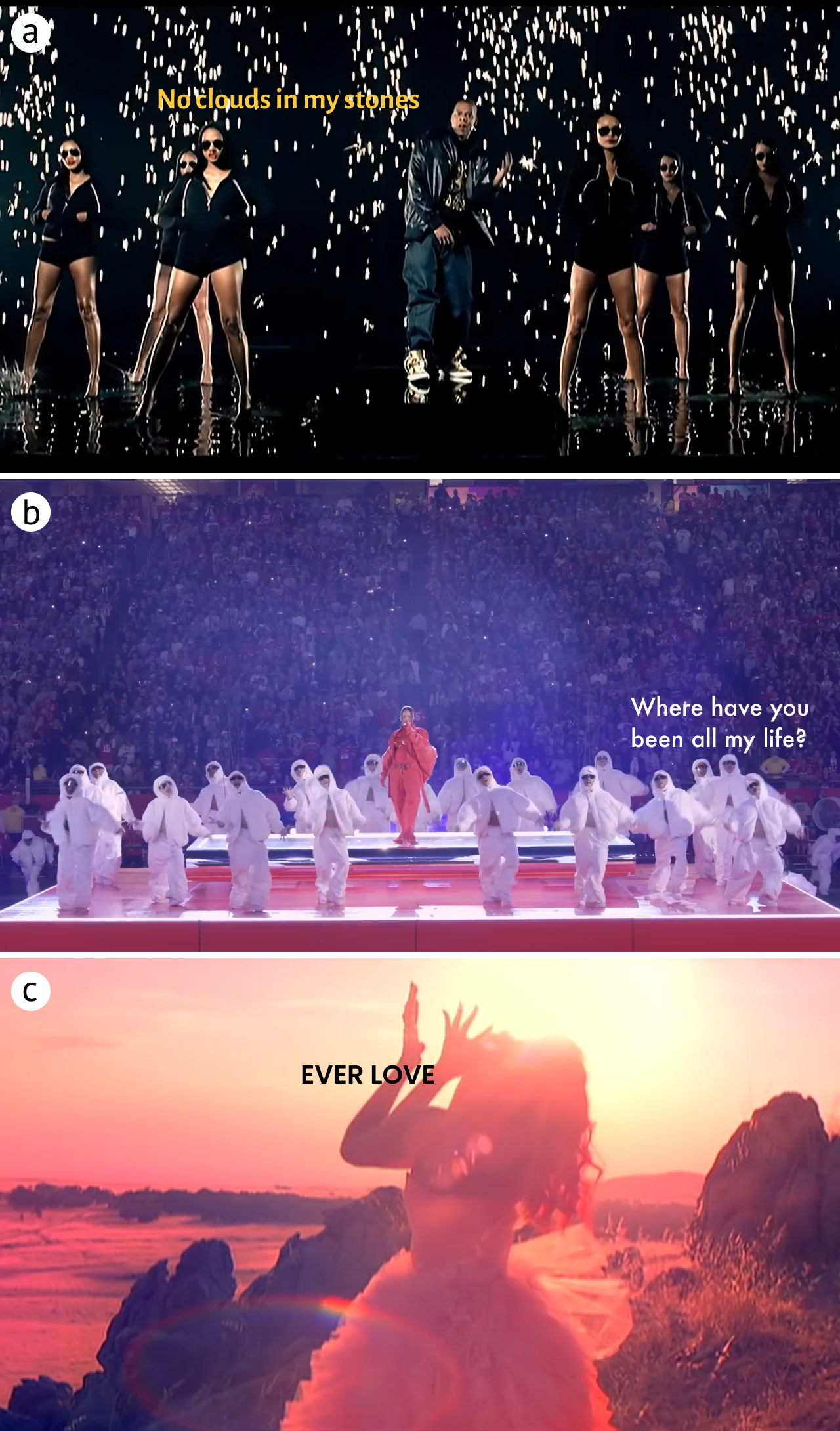}
    \caption{
    \camready{
    We find three recurring failure cases when generating lyric videos with our pipeline.
    When the input video is visually cluttered, it is hard to place the text in an easy-to-read way (a).
    Our focus of attention algorithm might find the wrong focal person when many similarly looking people are present (b).
    Our video object segmentation model [24] does not work well when the foreground and background are not clearly separated (c).
    %
    \emph{Video sources:
    (a) Umbrella by Rihanna \textcopyright 2007 The Island Def Jam Music Group~\cite{Rihanna:2009:U},
    (b) Rihanna's FULL Apple Music Super Bowl LVII Halftime Show~\cite{NFL:2023:HS}, 
    (c) Only Girl (In The World) by Rihanna \textcopyright 2010 The Island Def Jam Music Group~\cite{Rihanna:2010:OG}.
    }
    }
    }
    \label{fig:cases}
\end{figure}

\section{Failure Cases}
\label{sec:cases}

\camready{
We tested many videos with our automated pipeline and found 3 recurring failure cases (\Cref{fig:cases}).\\
\begin{description}[style=unboxed,leftmargin=0cm]
    \item[Case 1: High visual clutter]
    If the input video is visually cluttered, our pipeline might not place the text in an easy-to-read way because it is difficult to ensure good contrast.
    This is shown in \Cref{fig:cases}a, where bright raindrops appear against a dark background.
    As discussed in Section~\ref{sec:limitations}, one possible fix in the future is to make a region of the background blurred or in solid color.\\

    \item[Case 2: Focus of attention algorithm limitations]
    Our algorithm looks for people or objects referenced by the lyrics.
    When they cannot be detected, the algorithm outputs empty masks that can result in text placed far from the focus of attention (Section~\ref{subsubsec:focus}).
    Further, we define the person who appeared most frequently as the foreground object.
    We may find the wrong focal person when the correct one is surrounded by many similarly looking people, like the dancers in \Cref{fig:cases}b.
    Currently, the user can fix these by drawing a bounding box around the desired foreground object in a shot's first frame.\\

    \item[Case 3: Tracking model limitations]
    As discussed in Section~\ref{sec:limitations}, our video object segmentation model~\cite{Paul:2022:RTS} does not work well on some videos.
    The video shown in \Cref{fig:cases}c has a pink tone that mixes the foreground and background, while the foreground object also has fast motions.
    The model cannot consistently track the foreground object.
    This results in text placements that occlude the foreground object.
    Currently, this can be manually fixed by tracking models that supports interactive editing~\cite{Cheng:2022:XMem}.
\end{description}
}

\section{Videos Used in User Evaluation}
\label{sec:list}
\camready{
\begin{enumerate}
    \item I Ain't Worried by OneRepublic~\cite{OneRepublic:2022:IAW}
    \item We Don't Talk Anymore by Charlie Puth \& Selena Gomez~\cite{Puth:2016:WDTA}
    \item Ice Cream by BLACKPINK \& Selena Gomez~\cite{Blackpink:2021:IC}
    \item Selfish Soul by Sudan Archives~\cite{Sudan:2022:SS}
    \item August by Taylor Swift~\cite{swift:2021:A}
    \item Nervous by John Legend~\cite{Legend:2022:N}
    \item Lilacs by Waxahatchee~\cite{Waxahatchee:2020:L}
    \item Fire by Waxahatchee~\cite{Waxahatchee:2020:F}
    \item Flower by Miley Cyrus~\cite{Cyrus:2023:F}
    \item Let It Go by Idina Menzel~\cite{Menzel:2013:LIG}
\end{enumerate}
}

\section{User Evaluation Statistics}
\begin{table}[h]
    \begin{tabular}{llll}
                                           &                                                                                    &                                                                                     &                                                               \\
    \multicolumn{1}{l|}{Q1}                & \multicolumn{1}{l|}{Baseline}                                                      & \multicolumn{1}{l|}{Readability Ablated}                                            & Attention Ablated                                             \\ \hline
    \multicolumn{1}{l|}{Full Pipeline} & \multicolumn{1}{l|}{\begin{tabular}[c]{@{}l@{}}z = 277.5\\ p = 0.043\end{tabular}} & \multicolumn{1}{l|}{\begin{tabular}[c]{@{}l@{}}z = 121.5\\ p = 0.0004\end{tabular}} & \begin{tabular}[c]{@{}l@{}}z = 88.5\\ p = 0.0003\end{tabular} \\
                                           &                                                                                    &                                                                                     &                                                     \\
    \multicolumn{1}{l|}{Q2}                & \multicolumn{1}{l|}{Baseline}                                                      & \multicolumn{1}{l|}{Readability Ablated}                                            & Attention Ablated                                             \\ \hline
    \multicolumn{1}{l|}{Full Pipeline} & \multicolumn{1}{l|}{\begin{tabular}[c]{@{}l@{}}z = 254.5\\ p = 0.034\end{tabular}} & \multicolumn{1}{l|}{\begin{tabular}[c]{@{}l@{}}z = 222\\ p = 0.001\end{tabular}}    & \begin{tabular}[c]{@{}l@{}}z = 96\\ p = 0.000006\end{tabular} \\
                                           &                                                                                    &                                                                                     &                                                               \\
    \multicolumn{1}{l|}{Q3}                & \multicolumn{1}{l|}{Baseline}                                                      & \multicolumn{1}{l|}{Readability Ablated}                                            & Attention Ablated                                             \\ \hline
    \multicolumn{1}{l|}{Full Pipeline} & \multicolumn{1}{l|}{\begin{tabular}[c]{@{}l@{}}z = 309\\ p = 0.164\end{tabular}}   & \multicolumn{1}{l|}{\begin{tabular}[c]{@{}l@{}}z = 235\\ p = 0.003\end{tabular}}    & \begin{tabular}[c]{@{}l@{}}z = 150\\ p = 0.0006\end{tabular}
    \end{tabular}
    \caption{Statistics of the post-hoc pairwise Wilocoxon test.}
    \label{table:stats}
\end{table}

%

\end{document}